# Multimode Objective Lens for Momentum Microscopy and XPEEM: Experiments

O. Tkach[1,2,*], S. Fragkos[3], D. Biswas[4], J. Liu[4], O. Fedchenko[1,5], Y. Lytvynenko[1,6], S. Babenkov[1], D. Zimmer[1], Q. Nguyen[7], S. Chernov[8], D. Kutnyakhov[8], M. Scholz[8], N. Wind[8,9], A. Gloskowskii[8], F. Pressacco[8], J. Dilling[10], L. Bruckmeier[10], M. Heber[8], L. Wenthaus[8], G. Brenner[8], D. Puntel[7], P. E. Majchrzak[7], D. Liu[7], F. Scholz[8], J. A. Sobota[7], J. D. Koralek[7], G. Dakovski[7], A. Mehta[7], N. Sirica[11], M. Hoesch[8], C. Schlueter[8], L. V. Odnodvorets[2], Y. Mairesse[3], T.-L. Lee[4], A. Kunin[12], K. Rossnagel[10,13], Z. X. Shen[14], H.-J. Elmers[1], S. Beaulieu[3], and G. Schönhense[1*]

***1*** *Johannes Gutenberg-Universität, Institut für Physik, D-55099 Mainz, Germany*
***2*** *Sumy State University, Kharkivska Str. 116, 40007 Sumy, Ukraine*
***3*** *Université de Bordeaux - CNRS - CEA, CELIA, UMR5107, F33405 Talence, France*
***4*** *Diamond Light Source Ltd., Didcot OX11 0DE, United Kingdom*
**5** *Physikalisches Institut, Goethe-Universität Frankfurt, Campus Riedberg, Max-von-Laue-Str. 1, 60438 Frankfurt am Main, Germany*
***6*** *V.G. Baryakhtar Institute of Magnetism of the NAS of Ukraine, 03142 Kyiv, Ukraine*
***7*** *SLAC National Acceleration Laboratory, Menlo Park, CA 94025, USA*
***8*** *Deutsches Elektronen-Synchrotron DESY, D-22607 Hamburg, Germany*
***9*** *University of Hamburg, Institut für Experimentalphysik, D-22761 Hamburg, Germany*
***10*** *Institut für Experimentelle und Angewandte Physik, Christian-Albrechts-Universität zu Kiel, D-24098 Kiel, Germany*
***11*** *Los Alamos National Laboratory, Santa Fe, NM 87545, USA*
***12*** *Princeton University, Department of Chemistry, Princeton, NJ 08544, USA*
***13*** *Ruprecht Haensel Laboratory, Deutsches Elektronen-Synchrotron DESY, D-22607 Hamburg, Germany*
***14*** *Stanford University, Stanford, CA 94305, USA*

** Corresponding authors: tkachole@uni-mainz.de; schoenhe@uni-mainz.de*

**Abstract**

A new type of objective lens has recently been proposed for use in X-ray photoemission electron microscopes (XPEEMs) and momentum microscopes. Adding a ring electrode concentric with the extractor allows the field in the gap between the sample and the extractor to be shaped. Forming a lens field in this gap reduces the field strength at the sample by up to an order of magnitude. This mitigates the risk of field emission, particularly for cleaved samples with sharp edges. A retarding field can redirect all slow electrons, thus eliminating the primary contribution to the space-charge interaction. Here we present the first experimental investigation of the new lens, examining its performance at photon energies ranging from the extreme ultraviolet produced by a high-harmonic generation (HHG)-based source to soft and hard X-rays at two synchrotron facilities. The gap lens in a region without electrodes enables large working distances up to 23 mm. Reduced aberrations allow for larger fields of view in both *k*-space and real-space imaging, with resolutions comparable to those of conventional cathode lenses. However, field strengths are an order of magnitude smaller. The zero-field mode enables the study of 3D structured objects and is therefore beneficial for small cleaved samples as well as for *operando* devices involving top electrodes. The repeller mode reduces space-charge effects, but results in a smaller *k*-field diameter. This reduction ranges from 10% at hard X-ray energies to 50% in the XUV range. The usable energy interval is also reduced by a factor of two. In time-of-flight XPEEM mode the raw data show a resolution of 250 nm, which can be improved to better than 100 nm through data processing.

## 1. Introduction

Using the cathode-lens approach, a photoelectron momentum microscope (MM) can simultaneously record *k*-patterns over large intervals in energy-momentum parameter space. The core of a cathode lens is the high accelerating field between the sample and the extractor. Cathode lenses were originally developed for low-energy electron microscopy (LEEM) [1,2] and photoemission electron microscopy (PEEM) [3,4]. Various energy filters have been developed for MMs. An early design used a simple retarding-field energy filter [5]. Later designs used a tandem configuration of two hemispherical analyzers [6-8] or a single hemispherical analyzer [9-12]. Unprecedented efficiency was achieved through time-of-flight (ToF) energy recording, which enables the simultaneous acquisition of three-dimensional $I(k_x,k_y,E_{kin})$ data files [13] and can be combined with a hemispherical analyzer [14].

The high degree of parallelization in data recording makes MMs particularly useful for low-intensity experiments. Examples include angle-resolved photoelectron spectroscopy (ARPES) in the extreme ultraviolet (XUV) range [15, 16], the soft X-ray range [17, 18], and the hard X-ray range [19]. One rapidly growing field is ultrafast time-resolved photoelectron spectroscopy using femtosecond photon pulses. Experiments using the pump-probe approach require adjusting several parameters and thus demand a recording technique with a high degree of parallelization. In just five years, MM experiments in the ultrafast regime have established a new frontier of research conducted by a few groups. See Sec. 2.1 for details. MM avoids the effects of sample or analyzer rotation. Thus, it is ideal for recording circular and linear dichroism in photoemission and for combining it with the imaging spin filter approach.

In this article we discuss the advantages and limitations of the cathode-lens approach. The high electric field can cause field emission or flashovers at spots with locally enhanced field strength, such as the edges of cleaved samples. Additionally, the extractor field effectively collects ‘unwanted’ secondary electrons related to the true photoelectron signal, as well as electrons generated by multiphoton photoemission (nPPE) through pump pulses in the visible or infrared spectrum. These low-energy electrons are pulled into the lens column, where they repel the photoelectrons via Coulomb interaction. This results in space-charge shifts and broadening [20]. Furthermore, the slow electrons can lead to detector saturation.

We present here the first set of measurements taken with a new objective lens. This lens replaces the homogeneous extractor field with specially shaped electric fields within the gap between the sample and the extractor. One or more ring electrodes, which are concentric with the extractor, generate various types of lens fields. Varying the potentials of the ring electrodes adjusts the field strength at the sample surface over a wide range, from accelerating to retarding. The ‘gap lens’ is formed in an area without electrodes. This allows for larger working distances between the sample and the first set of electrodes. The new lens design was recently studied using ray tracing and aberration coefficient calculations [21]. The present experimental study of the different lens modes (extractor, gaplens, zero-field, and repeller mode) confirms the theoretical predictions. Measurements were performed using various photon sources, including high-harmonic generation (HHG)-based XUV and soft and hard X-ray energies, at two synchrotron sources: PETRA III (DESY, Hamburg) and the Diamond Light Source (Didcot, UK).

## 2. Motivation for the New Objective Lens

### *2.1 Strengths and challenges of momentum microscopy*

Momentum microscopy is an emerging technique that can simultaneously capture a large ($k_x$,$k_y$,$E_{kin}$) parameter space, which is its key advantage. This high recording efficiency is especially important for low-intensity experiments, such as time-resolved ARPES. Accessing the full $2\pi$ solid angle without rotating the sample rules out modulation by variation of the matrix element due to a change in the angle of incidence. Therefore, MM is ideal for studying circular and linear dichroism in photoemission. Classical ARPES experiments with a deflector lens in front of a hemispherical analyzer can also record 3D photoelectron distributions. However, this approach only observes patterns within a limited range of polar angles. Thus, the accessible *k*-range is limited at low kinetic energies unless the sample is rotated. Dichroic ToF-MM experiments at synchrotron sources [13,22-24] have recently been complemented by the use of circularly polarized HHG-based XUV light [25,26]. Finally, the 2D or 3D recording architecture of MMs is well-suited for combination with imaging spin filters [7,8,15-17,27,28].

The large phase-space acceptance with fixed geometry, even at low kinetic energies, has proven to be extremely powerful for time-resolved pump-probe experiments on nonequilibrium dynamics. This is because it simultaneously captures the dynamics of hot electrons and holes with energies up to several eV above and below the Fermi level ($E_F$). Since there are no review articles on this topic yet, we will briefly summarize some recent benchmark experiments to which ToF-MM contributed substantially.

Early experiments were performed using the 'HEXTOF' [20], a ToF-MM at the PG2 soft X-ray beamline at FLASH. Dendzik *et al*. used higher photon energies to precisely observe the correlation between core-level and conduction-band dynamics [29]. This was followed by real-time observation of the metamagnetic phase transition in FeRh [30], the first ultrafast core-level and photoelectron diffraction experiments [31,32], and the capture of electron-phonon energy flow in laser-heated nickel [33]. An emerging application of ToF-MM is ultrafast orbital tomography. The first experiments, performed at FLASH [34] and using HHG sources [35,36], suggest that time-resolved tomography of molecular wave functions during chemical reactions will soon be possible.

Several groups in Germany, Japan, France, and the USA perform HHG-based ToF-MM experiments. These experiments focus on the dynamics and intrinsic timescales of phase transitions in materials [37,38], exciton formation, and retrieving the corresponding excitonic wavefunction [39-44], including Moiré interlayer excitons in layered materials [45-47]. In addition to providing deep insight into ultrafast dynamics, ToF-MM is well-suited for studying time-resolved dichroism phenomena in photoelectron patterns. This has been demonstrated for time-reversal and Fourier dichroism [48,49]. Theory predicts Floquet physics and associated light-induced topological phenomena [50]. Floquet physics using conventional time-resolved ARPES (6 eV probe) was pioneered in Ref. [51]. The first time-resolved ToF-MM experiments of Floquet states were conducted with 2H-$WSe_2$ [26] and graphene [52] .

The PETRA III synchrotron beamlines P04 and P22 host specialized ToF-MMs that can accommodate higher photon energies. Recent relevant papers have focused on the rearrangement of electronic states upon Néel vector switching in the collinear metallic antiferromagnet $Mn_2Au$ [53] and on time-reversal symmetry breaking in the electronic structure of the altermagnet $RuO_2$ [23]. A laboratory experiment using a Ti:sapphire laser studied plasmonic emission from resonantly excited nanoparticles (see [54] and references therein). Examples of spin-filtered ToF-MM are shown in the review [55] and the application of full-field photoelectron diffraction is discussed in [56]. The most recent achievement was the implementation of a hybrid hemisphere-ToF MM at beamline I09 of the Diamond Light Source [14]. Over 60 publications in the last three years demonstrate that ToF-MM is becoming a well-established method. However, challenges remain regarding the undesired effects of the extractor field.

The most severe challenge in experiments with conventional cathode objective lenses is caused by the high electric field between the sample and the extractor. Fields of 3 – 8 kV/mm typically pose a risk of field emission or flashovers at spots with a locally enhanced electric field. These spots can appear on cleaved samples with sharp corners, on layered samples with protruding lamellae, and on samples with other microscopic defects. The likelihood of these phenomena often exhibits threshold behavior. The conventional approach to addressing this issue is to reduce the extractor voltage, which compromises resolution [3,4].

A second challenge is the space charge interaction, which occurs in experiments with pulsed radiation. Intense pump-laser pulses can photoemit a large number of slow electrons, particulary from 'hot spots', as was first observed in Ref. [57]. Nanoscale objects and defects can act as plasmonic resonators, amplifying the optical near-field by orders of magnitude, as observed in PEEM [58]. These resonators can emit electrons through various mechanisms, including near-field-enhanced multi-photon photoemission (nPPE), optical field emission, and the single-particle decay of localized surface plasmon-polaritons [59]. Additionally, secondary electrons originating from cascade-like inelastic scattering processes of fast photoelectrons contribute to space charge effects. The number of these electrons increases with increasing photoelectron kinetic energy. Space-charge-induced shifts of up to 10 eV have been observed in valence-band photoemission from Ir(111) using 50 ps synchrotron pulses at $h\nu$= 1000 eV [60]. A semiempirical model revealed a total charge of the slow-electron cloud that exceeds the number of true valence-band photoelectrons by four to five orders of magnitude (see Figs. 1 and 2 in Ref. [60]). A particularly serious scenario occurs when the pulse intensity fluctuates, as with SASE-type FELs. In this case, the shift varies from pulse to pulse, resulting in space-charge broadening of the signal.

In summary, the ToF-MM method significantly benefits time-resolved pump-probe photoemission experiments. This method acquires $k$-patterns over large ($k_x$,$k_y$,$E_{kin}$) intervals without moving or rotating the sample or deflecting the beam. Using the PEEM mode, we can identify homogeneous regions on the sample and optimize the spatial and temporal overlap of the pump and probe beams [20]. However, the strong electrostatic field of the cathode lens can cause field emission or flashovers and pulls all slow electrons into the lens column, where they cause space charge effects. These challenges are connected to the extractor field and require alternative experimental solutions, as demonstrated here.

### *2.2 Operating principle of the multimode lens*

The study of space-charge effects within a semi-empirical model [60] led to the design of specialized lens arrangements [61], which create a retarding field at the sample surface. Redirecting the slow electrons back to the sample significantly reduces their contribution to the Coulomb interaction in the beam. The first experiments at FLASH employed standard cathode lens optics and set the extractor electrode to a negative voltage with respect to the sample. These results confirmed the predictions and revealed complete suppression of the artifacts caused by space charge interaction [62]. Recently, various objective lens geometries with an additional electrode were analyzed using ray tracing and aberration coefficient calculations [21]. This study resulted in an electrode geometry that enables several operating modes through 'field shaping' of the gap region between the sample and the extractor. Here, we present the first experimental study of the new lens design.

Figures 1(a-d) show four front-lens modes, along with the calculated equipotential contours in the gap region between the sample (Sa) and the extractor (Ex). To make the front part visible, the lens photo is shown in perspective view. The key element is the ring electrode (R), which is concentric with the extractor electrode. The potential of the ring electrode allows the field in the gap to be tailored, as illustrated by the field contours. The downstream lens voltages need to be adjusted when switching between the different modes. For more details about the complete lens system, see Fig. 1 in Ref. [21]. A high positive potential on Ex and R generates a strongly accelerating homogeneous field. This is the classical *extractor mode* [see Fig. 1(a)]. We show two examples of photoelectron diffraction (PED) patterns of a Ge(001) sample recorded using the MM at the DIAMOND light source. The patterns were captured at the 3*d* core level with final state energies (i.e., kinetic energies within the solid) of 106 and 172 eV, as shown in Figs. 1(e) and (f), respectively. The large, planar Ge crystal enabled the extractor mode with $U_{Ex} = U_R = 12$ kV, producing a field of 3 kV/mm within the 4 mm gap.

Setting the voltage of lens element R to a lower value creates an additional converging lens (gap lens) between Sa and Ex, as illustrated in Fig. 1(b). Depending on the voltages chosen at Ex and R, the *gaplens mode* yields a significantly smaller accelerating field at the sample surface. We show band mapping results at two different photon energies as examples. The patterns in Figs. 1(g) and (h) were recorded at the HHG-based XUV source ($h\nu = 21.6$ eV) at CELIA in Bordeaux and at the soft X-ray beamline P04 ($h\nu = 440$ eV) of the PETRA-III synchrotron, respectively. In the CELIA experiment, the field was 600 V/mm, which is five times smaller than in the cases shown in Figs. 1(e, f). See sections 3.2 and 3.5 for more details. The additional lens in the gap allows for a significantly larger distance between the sample and the first electrode than in conventional extractor optics. This is advantageous when placing elements for photon beam focusing or gas dosing close to the sample.

By applying a negative voltage at R, we can tune the field at the sample surface to zero (*zero-field mode*), as shown in Fig. 1(c). This is advantageous for samples with topside electrodes or activators for operando studies. Low-energy electrons are overfocused, reducing the number of unwanted background electrons pulled into the lens column. Furthermore, a zero field enables measurements on tilted samples with large off-normal emission angles at high energies. The equipotential plots in Figs. 1(b) and (c) are quite similar. This suggests that the *zero-field mode* can be considered a *gaplens mode* with the special property of having a vanishing electric field strength at the sample surface. The figures show PED patterns recorded at the hard X-ray beamline P22 ($h\nu = 6$ keV) of PETRA-III. Fig. 1(i) shows the PED diffractogram of the oxygen 1*s* core level in $SrTiO_3$, which was recorded with the sample tilted 60°. We observed the influence of the phase transition on the PED pattern; for details, see Ref. [56]. Fig. 1(j) shows the PED diffractogram of the 2*s* core level for a Si(001) sample with a 0° tilt. Reducing slow electrons significantly improves resolution and contrast as well as the signal-to-noise ratio (see Sec. 3.1 for further details).

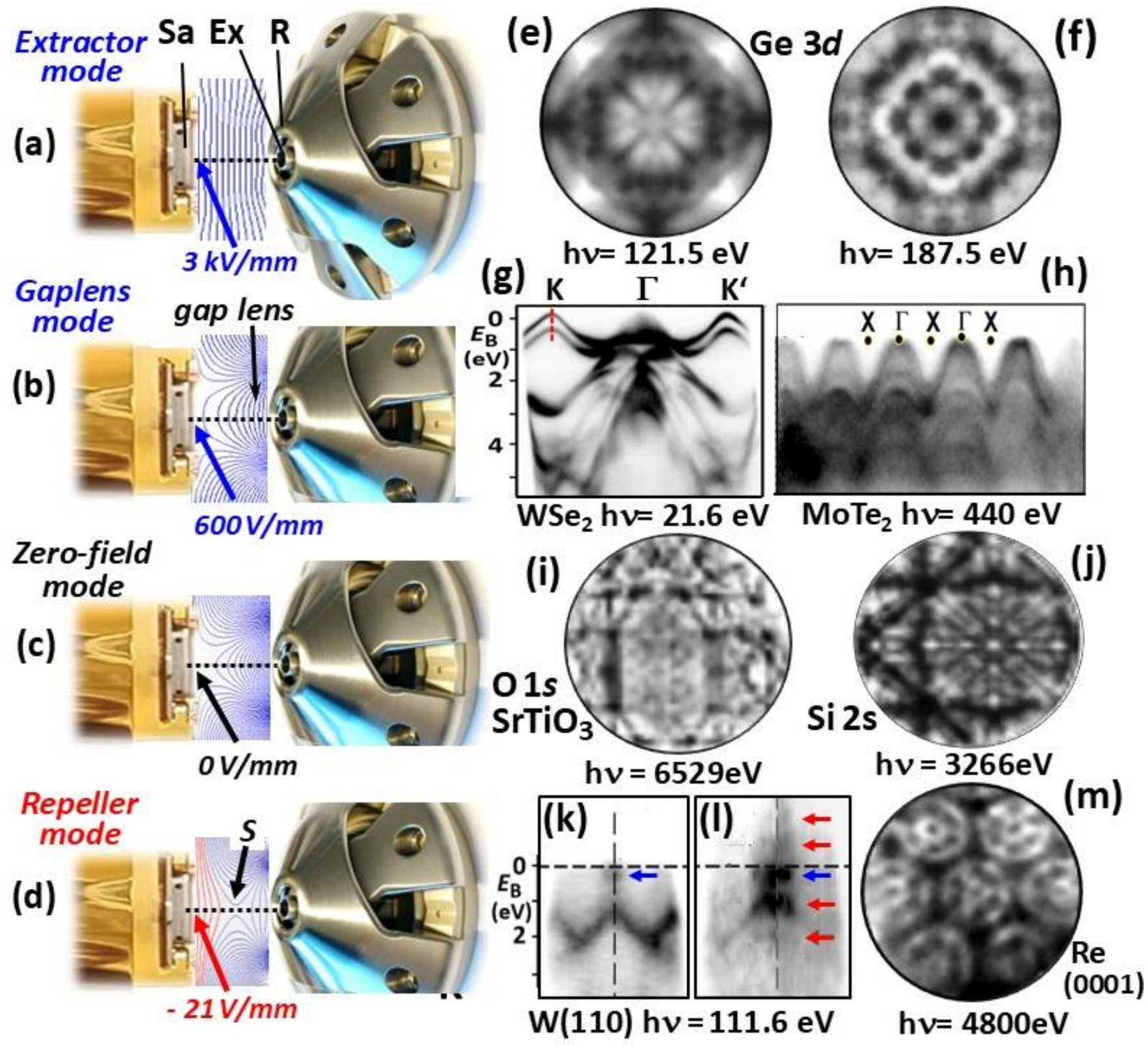

**Fig. 1.** Operation modes of the multimode lens, illustrated by equipotential contours for four different field configurations. (a) The classical *extractor mode* with extractor (Ex) and ring electrode (R) on high positive potential (here 12 kV) with respect to the sample (Sa), generating a homogeneous field of 3 kV/mm in the gap. The photo is shown in a perspective to reveal the front of the electrodes. (b) *Gaplens mode* with additional lens in front of the extractor. Here $U_R$ = 0 V, which reduces the field at the sample to 600 V/mm. (c) The *zero-field mode*, in which the fields of extractor and ring electrodes cancel each other out at the sample surface. (d) *Repeller mode* with a retarding field at the sample (red equipotentials) and a saddle point (S) at $U_S$ = -50 V, defining the low-energy cutoff. (e-m) Examples measured for the different modes (s. text). Data for (e,f) from [14], (g) see Sec. 3.5, (i) from [56], (j) from [63], (k,l) from [20], and (m) from [64].

Finally, a larger negative voltage at R and a reduced voltage at Ex generate a retarding field at the sample, resulting in the *repeller mode* (Fig. 1(d)). Red potential contours mark the retarding field. This mode is crucial for femtosecond pump-probe experiments. The valence band example of W(110), shown in Figs. 1(k,l), was measured at the free-electron laser FLASH (hν= 111.6 eV). In this case, the field at the sample was -21 V/mm, redirecting the slowest electrons up to $E_{kin}$= 5 eV back to the sample within the first 240 μm. The saddle point *S* acts as a high-pass filter in the potential landscape, here with $U_S$ = -50 V with respect to the sample potential. In *zero-field mode* (Fig. 1(c)), the saddle point shifts to the sample surface, thus defining the onset of the *repeller mode*. Panel (k) shows the tungsten valence band in the probe-only case. Panel (l) was captured with an additional pump pulse at zero pump-probe delay (hν = 1.2 eV; pump fluence 13 mJ/cm$^2$). The red arrows in panel (l) indicate the laser-assisted photoemission (LAPE) sidebands. For more information, see Ref. [62]. The repeller mode eliminates most of the Coulomb forces between the slow and fast electrons in the lens column, yielding higher resolution in static experiments as well. The Re(0001) valence band example, shown in Fig. 1(m), was measured at beamline P22 (PETRA-III).

# 3. Performance of the Multimode Lens in Momentum-Space Imaging

### *3.1 ToF MM in the hard X-ray range*

For low initial kinetic energies, which are typical for PEEM or LEEM experiments, the extractor field generates a virtual image of the sample with high kinetic energy, $E_{kin}+eU_{Ex}$, and a small angular divergence α'. $U_{Ex}$ is defined with respect to the sample potential. This virtual image forms at approximately twice the gap distance from the extractor (see Fig. 2 in Ref. [3]).

$$\alpha' = \sqrt{[E_{kin}/(E_{kin} + eU_{Ex})]}\alpha_0 \qquad (1),$$

where $E_{kin}$ and $\alpha_0$ are the initial photoelectron kinetic energy and starting angle on the sample, respectively. However, in the hard X-ray range $E_{kin}$ is typically several keV (e.g., up to 7 keV in [65]). At such energies, the extractor field's contribution to the total refractive power of the front lens is quite weak, and the virtual image model is invalid. The square root factor in front of $\alpha_0$, also called the electron-optical 'immersion ratio', is of the order of 1. On the other hand, the angular range of interest becomes very small. At 7 keV, a *k*-field diameter of 12 Å$^{-1}$ (consisting of many Brillouin zones [64]) corresponds to a polar angular range of only +/- 8°.

We studied the influence of the extractor voltage for photoelectron kinetic energies ranging from 1.43 to 5.88 keV. Figure 2 shows the results of measurements taken with the momentum microscope located at beamline P22 of PETRA III, which has an integral dodecapole bandpass filter for background suppression [62]. This instrument has a classical front lens without a repeller electrode, so it cannot be operated in the modes discussed in Fig. 1. The top row displays the Ge 3*p* photoelectron diffraction patterns at $E_{kin}$ = 3.16 keV (a-c) and 5.88 keV (d); the bottom row displays the same for Si 1*s* at $E_{kin}$ = 1.43 keV (e-g) and 4.16 keV (h). The first column shows the patterns with the extractor 'on' ($U_{Ex}$ = 8 kV), and the second shows patterns with the extractor 'off' ($U_{Ex}$ = $U_{Sa}$). Unlike the potential distribution shown in Fig. 1(c), here the entire space between sample and extractor is field free and hence there is no lens action. The third column shows the patterns with the extractor at a negative potential with respect to the sample, generating a uniform retarding field without a saddle point.

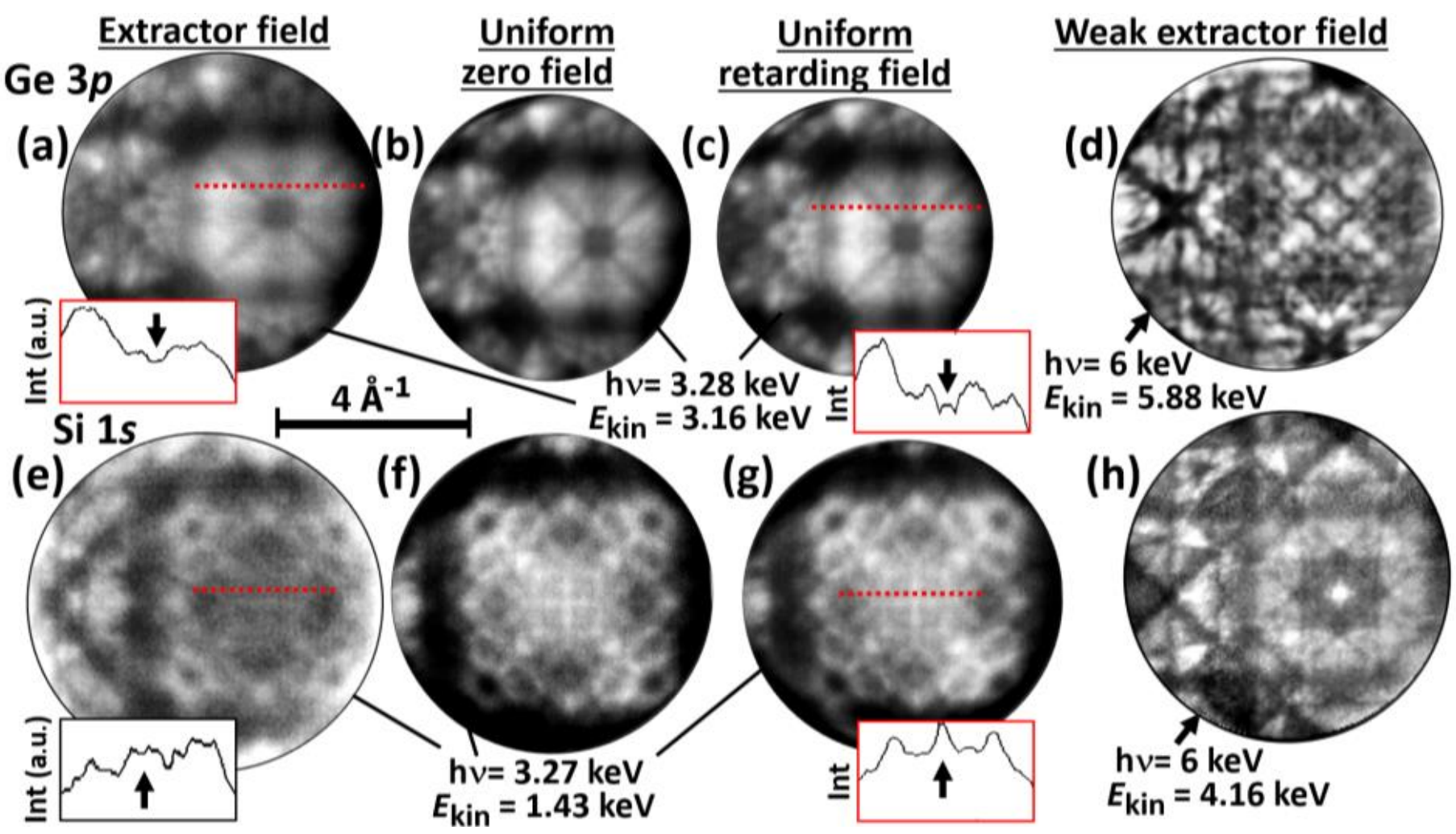

**FIG. 2.** Hard X-ray photoelectron diffraction (hXPD) of core-level electrons. (a,b,c) Ge 3*p* patterns at $E_{kin}$= 3.16 keV recorded with $U_{Ex}$ = 8 kV, $U_{Ex}$ = 0 V and $U_{Ex}$ = - 650 V, respectively. (d) The same as (a), but for $E_{kin}$ = 5.88 keV and $U_{Ex}$ = 4.1 kV. (e,f,g) Si 1*s* hXPD patterns at $E_{kin}$ = 1.43 keV, recorded with $U_{Ex}$ = 8 kV, $U_{Ex}$ = 0 and $U_{Ex}$= - 220 V, respectively. (h) Same as (d) but for Si 1*s* and $E_{kin}$ = 4.16 keV. The positive sample bias was $U_{Sa}$ = $E_{kin}$/e. The insets show intensity line scans along the dotted red lines. The scale bar is valid for all panels.

Comparing image sequences in Figs. 2(a,b,c) and (e,f,g) reveals an increase in image contrast and resolution. This increase can be attributed to the decrease in space-charge interaction and is most evident in the second row at a low kinetic energy of 1.43 keV, see Figs. 2(e) and (g). The effect is visible even with a *uniform zero field* because far fewer slow electrons enter the lens column without a field. With a *uniform retarding field*, all slow electrons are eliminated from the beam at a short distance from the sample surface. The retarding field in Fig. 2(g) is -220 V/mm. Thus, electrons with kinetic energies below 220 eV cannot enter the column, and the slowest secondary electrons ($E_{kin}$< 10eV) are redirected back within the first 50 µm above the sample surface. This means that the Coulomb interaction between the photoelectrons and the large number of slow electrons is 'switched off' in the initial phase following the photon pulse. The space-charge blur is due to the intrinsic, Lorentzian-shaped pushing action of the slow electrons [60]. This effect is visible in the intensity line scans. In Fig. 2(c), the dark vertical double line is resolved (see arrow in the inset), whereas it appears blurred in Fig. 2(a). In Fig. 2(g), the bright '+' in the image center appears as a sharp, bright peak (arrow in the inset), while this feature is practically invisible in Fig. 2(e).

Along with the increase in image contrast, we observe a reduction in the *k*-field diameter when switching from the *extractor field* (first column) to the *uniform zero field* and *retarding field* (2$^{nd}$ and 3$^{rd}$ columns). For $E_{kin}$ = 3.16 keV, the reduction is 20% when changing from extractor to zero or retarding fields (a-c). For the lower initial energy of $E_{kin}$ = 1.43 keV (bottom row), the reduction is 30%. Due to the small retarding voltages of $U_{Ex}$=-220 and -650V, the difference in k-field between zero and retarding fields is negligible, cf. Figs. 2(b,c) and (f,g). At higher kinetic energies of 5.88 and 4.16 keV, Figs. 2(d,h), the difference between the uniform zero field and the weak extractor field is also negligible, even with $U_{Ex}$ = 10 kV.

From these measurements using a standard objective lens, we conclude that the slow electrons contribute significantly to space-charge effects when the extractor field is on, even in synchrotron experiments at high energies with rather long photon pulses (here, 70 ps). This influence is reduced with uniform zero field because only electrons within a small solid angle interval can enter the microscope through the small extractor hole. By limiting the acceptance bandwidth of the front lens by a weak retarding field, the space charge in the column can be suppressed. A uniform zero field has been used previously for measurements with tilted samples at large off-normal angles (see Fig. 2(c) in [64] and Figs. 5(j-p) in [56]), while a weak extractor field was used for graphite measurements in [65]. Space-charge effects have also been identified as a resolution-limiting factor in XPEEM. For more information, see Sec. 4.1. The energy widths of the core level signals studied in Fig. 2 are about 0.5 eV. For this width, the chromatic aberration of the front lens is negligible (see Tab. 1 in Ref. [21]).

### *3.2 Time-of-flight MM in the soft X-ray range*

The soft X-ray regime comprises photon energies ranging from ~100 eV to 2 keV. The lower and upper limits border the XUV and the tender X-ray ranges, respectively. This energy range is non-trivial for simulations and electron optics performance, since the 'immersion ratio' in Eq. (1) varies significantly with kinetic energy. While the model of a virtual image at high energy with small $\alpha'$ (see Eq. (1)) remains valid at 100 eV, it becomes invalid as $E_{kin}$ increases. Tromp et al. have refined the theoretical treatment [2]. According to the simulations [21], the *k*-images recorded in the *gaplens mode* exhibit significantly lower spherical aberration than those recorded in *extractor mode*, despite the lower field at the sample surface. Here, we present the experimental validation of this prediction.

Figure 3 shows valence band results recorded at the upper and lower end of the soft X-ray range, comparing the *zero-field mode* with *extractor* and *gaplens modes*. Data for silicon carbide, Figs. 3(a-h), was taken at beamline P04 of PETRA III at hν= 1488 eV. Data for graphene, Figs. 3(i-o), was taken at the plane grating monochromator (PG2) beamline of the free-electron laser FLASH at hν= 83 eV. For SiC we show results for the uniform zero-field case (a,b,e,f) and *extractor mode* (c,d,g,h). During this initial measurement, the objective lens had no ring electrode, and the uniform zero-field was attained by adjusting extractor and sample to the same potential. The SiC band structure is well known; see, for example, ref. [66]. Here we present these measurements solely to discuss the quality of the *k*-patterns. The first row shows ($k_x$,$k_y$) and ($k_x$,$E_B$) sections through the $\overline{\Gamma}$-point along the $\overline{\Gamma}$ - $\overline{\mathrm{M}}$ direction. The second row shows analogous data along the $\overline{\mathrm{M}}$ - $\overline{\mathrm{M}}$ line. Figs. 3(b,f) and (c,g) correspond to different binding energies. We observe the prominent inner valence band head centred at the $\overline{\Gamma}$-point, which exhibits a circular shape with slight hexagonal warping, see panel (b). Panels (a) and (e) show sections along the dashed lines in (b) and (f), respectively. The band is nearly degenerate at $\overline{\Gamma}$, splitting into the $L_1$/$L_3$ bands along the $\overline{\Gamma}$ - $\overline{\mathrm{M}}$ direction. The $L_3$ band splits further into two bands along the $\overline{\mathrm{M}}$ - $\overline{\mathrm{M}}$ direction.

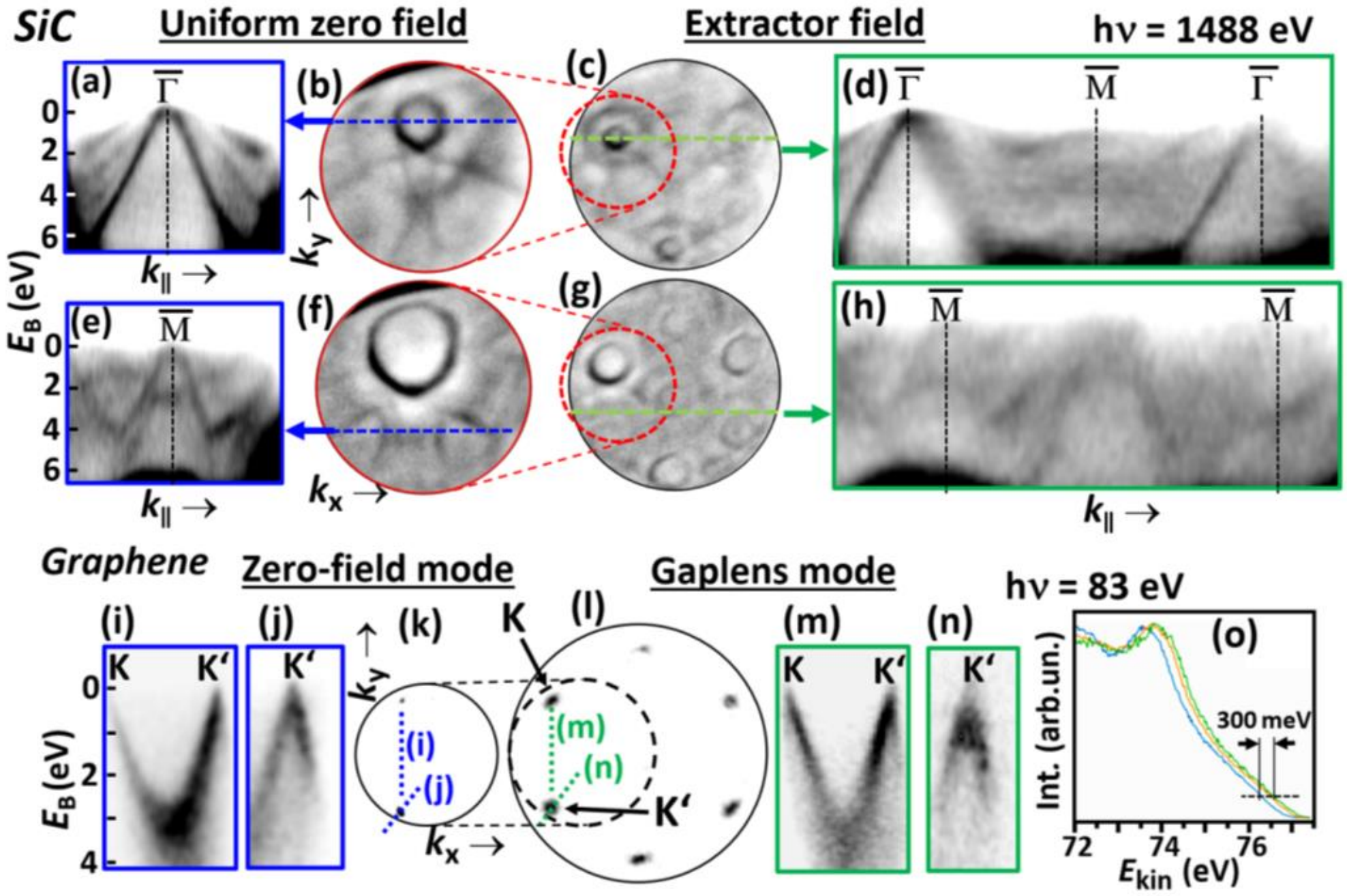

**FIG. 3.** (a-h) Valence-band patterns of SiC measured using the ToF-MM at the soft X-ray beamline P04 of PETRA III (hν = 1488 eV). (a,e) $k_x$-$E_B$ sections and (b,f) $k_x$-$k_y$ sections recorded using a *uniform zero-field*. The dashed blue lines in (b,f) indicate the positions of the sections displaying the band dispersions $E_B$-*vs*-$k_\parallel$ in (a,e). (d,h) and (c,g) Similar data, but recorded using the *extractor mode*. (i-n) Valence-band patterns of graphene, recorded at beamline PG2 of the free-electron laser FLASH (hν= 83 eV) using the *zero-field* and the *gaplens mode*. This data was obtained from a pump-probe experiment. In (i,j,k) and (l,m,n) the pump fluences were 3.10 mJ/cm$^2$ and 0.56 mJ/cm$^2$, respectively. (o) Corresponding *k*-integrated spectra for pump fluences of 3.10 mJ/cm$^2$, 2.25 mJ/cm$^2$ and 0.56 mJ/cm$^2$ (from left to right). The dashed circles in (c,g) and (l) denote the reduced *k*-fields in *zero-field mode*.

Figures 3(c,d) and (g,h) show the same sequence, which was measured with extractor on ($U_{Ex}$ = 10 kV). The features observed are the same, but appear more blurred than in the zero-field case*,* (a,b) and (e,f). However, when the extractor is turned on (c,g), the observed *k*-field is twice as large as it is in the uniform zero field (b,f), as indicated by the dashed circles in (c) and (g). Without a field between the sample and the extractor, there is no focusing, and the *k*-image is clipped at the hole in the extractor electrode.

The differences in contrast and resolution between the two modes are eye-catching in the first and second rows of Fig. 3, particularly in the $E_B$–*vs*–$k_\parallel$ sections (a,d,e,h). The bands appear much sharper and more contrasted in the patterns recorded with zero field (a,e) than with extractor on (d,h). Fig. 3(a) illustrates the splitting of the valence band into the three $L_1$ and $L_3$ bands as the distance from $\bar{\Gamma}$ increases [66].

The lower part of Fig. 3 shows results obtained from a graphene sample using femtosecond pulses at FLASH in a pump-probe scheme. These measurements were taken using the new front lens with a ring electrode. The sample was exposed to infrared laser pulses (pump, 1030 nm), which were synchronized with the FEL pulse train at hν = 83 eV (probe). The goal was to investigate the influence of pump intensity on the band patterns by comparing the *zero-field mode* ($U_{Ex}$ = 5.675 kV; $U_R$ = -3.1 kV) with the *gaplens mode* ($U_{Ex}$ = 11 kV; $U_R$ = 0 V). As with the SiC measurements, the contrast in the band patterns is greater in the *zero-field mode;* compare Figs. 3(i,j) and (m,n). The $k_x - k_y$ sections (Fig. 3(k,l)) demonstrate that the *k*-field diameter is also reduced by a factor of two in this case. Switching from the *gaplens* to the *zero-field mode* allowed the pump fluence to be increased from 0.563 mJ/cm$^2$ to 3.098 mJ/cm$^2$ without significantly degrading the *k*-patterns. Only a small, space-charge-induced (rigid) shift of 300 meV around zero pump-probe delay was visible; see the spectra in Fig. 3(o). For more details on ToF-MM experiments of graphene recorded with fs excitation, see Ref. [67].

The central question of this study was whether the excellent imaging properties predicted by ray tracing could be confirmed through experiments. The valence band results in Fig. 3 align with our findings in Fig. 2 regarding core levels, which show that the *zero-field mode* significantly improves the quality of the momentum patterns. The number of slow electrons that are pulled into the microscope is greatly diminished when the field at the sample is zero. Consequently, the blurring of the band features due to the Coulomb interaction between electrons is largely reduced and the pump fluence could be increased by a factor of five. The disadvantage of the *zero-field mode* is its smaller *k*-field with about half the size of the *k*-field in *extractor* and *gaplens mode*.

### *3.3 Hemisphere-based MM in the soft X-ray range*

The measurements in this section were performed using the single-hemisphere-based MM at beamline I09 of the Diamond Light Source. The soft X-ray branch houses a unique momentum microscope that combines a hemispherical analyzer with time-of-flight detection. This enables the simultaneous acquisition of $I(E_B,k_x,k_y)$ data arrays within the energy interval selected by the hemisphere [28]. The instrument was recently equipped with the multimode front lens. Results from the first beamtime with this lens are shown in Figs. 4 and 5. When the photon footprint is small, a ToF-MM does not require beam-confining apertures. However, a hemisphere-based MM requires entrance and exit apertures that define the energy resolution, see Eq. (1) in Ref. [28]. In *k*-imaging, a Gaussian image is focused on the entrance plane. The ray displacement $\delta_S$ due to the spherical aberration is defined by the coefficient $C_3$ and increases with the cube of the emission angle $\alpha_0$. For details, see Suppl. Material of [21].

$$\delta_S = C_3\, \alpha_0^3 \qquad (2)$$

Depending on its size, the entrance aperture may block the outer edges of the ray bundle, resulting in clipping of the *k*-field. Due to its lower aberrations, the *gaplens mode* produces a smaller beam crossover diameter at the entrance and exit apertures of the analyzer. This leads to larger observable *k*-fields than in the *extractor mode*. Using the $C_3$ values from Table 1 in [21], we can estimate the gain in the $\alpha_0$ range when the gap lens is activated. For electric field strengths between 880 and 1,200 V/mm the gain ranges from 1.6 to 1.7.

Figure 4 shows a series of measurements taken in *gaplens mode*. The top row displays valence band patterns of a Ge(001) sample recorded at a photon energy of 150 eV for pass energies ranging from 20 to 50 eV. At $E_{pass}$= 20eV [Fig. 4(a)] we zoomed into the 1st BZ with a k-field of 3 Å$^{-1}$. The distance of adjacent $\Gamma$-points is 2.216 Å$^{-1}$. At $E_{pass}$= 30 eV (b) and 40 eV (c) band features of the adjacent Brillouin zones become visible. Due to the strong curvature of the final state sphere (energy conservation), the repeated BZs are cut at different $k_z$ values, resulting in different shapes. At $E_{pass}$= 40 and 50 eV (c,d) the horizontal *k*-field size is 5 Å$^{-1}$, corresponding to an angular range of +/- 25°.

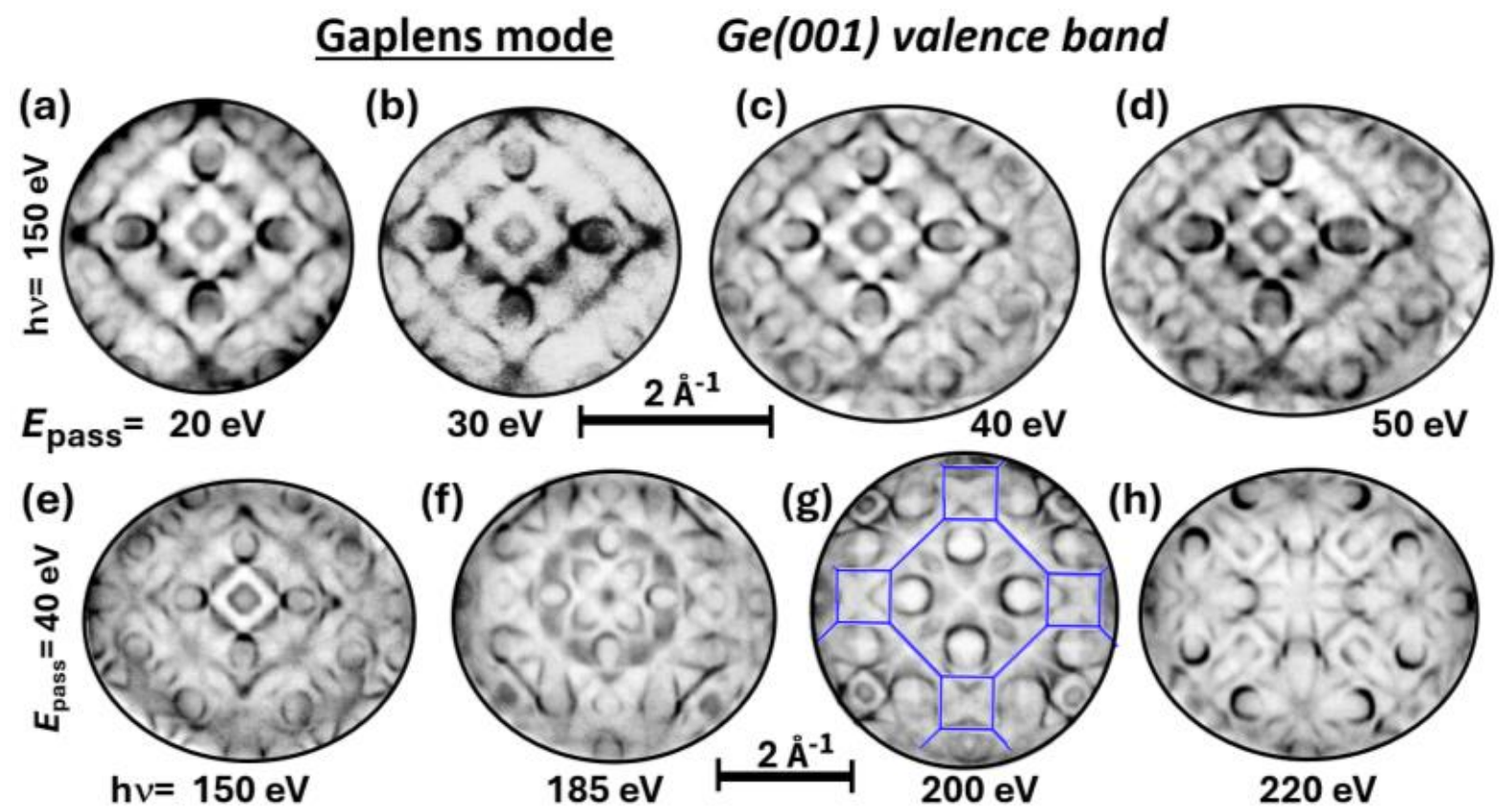

**FIG. 4.** Valence-band patterns of Ge(001) recorded in *gaplens* mode. For these measurements we used the hybrid hemisphere & ToF MM at beamline I09 at the Diamond Light Source in Didcot, UK. (a-d) Ge(001) measurements (h$\nu$ = 150 eV) at a binding energy of 3 eV and pass energies between 20 and 50 eV. The *k*-field diameter increases from 3 to 5 Å$^{-1}$. (e-h) Measurements at $E_{pass}$ = 40 eV for four different photon energies. A cut through the periodic pattern of BZs is sketched in (g), which provides the *k*-scale.

Figures 4(e-h) show a series of Ge(001) valence band patterns at $E_{pass}$ = 40 eV, recorded at various photon energies ranging from 150 to 220 eV. The BZ contour, blue lines in (g), provides a measure of the momentum scale, see the scale bars. The four patterns differ significantly because the final-state sphere intersects the periodic pattern of BZs at different $k_z$ values. The high contrast indicates good $k_z$ resolution. This suggests that Ge exhibits true bulk photoemission within this photon energy range.

All measurements were performed in *gaplens mode.* Despite the low field strength, the *k*-patterns are of good quality, confirming the simulations. We hypothesize that a key factor is the significantly reduced spherical aberration of the gap lens, as quantified by Eq. (2). The aberration coefficients $C_3$ given in Table 1 of Ref. [21], predict a gain in the emission angle range of 1.6 - 1.7 for electric field strengths of 880 - 1,200 V/mm.

### *3.4 Photoelectron diffraction*

First observed and discussed in the 1970s [68,69], photoelectron diffraction (PED) has become a powerful technique for analyzing the structure of solids. The full-field imaging MM approach has greatly improved the recording efficiency of PED patterns. Extending into the hard X-ray region provides access to true bulk information [55]. In particular, combining four operational modes is advantageous: (i) inspecting the sample in the PEEM mode quickly sets the photon spot to an optimum, defect-free position. (ii) XPEEM at selected core levels captures the distribution of elements, including their chemical state, in the probe volume (chemical structure; see Sec. 4.1 for details). (iii) Imaging PED enables rapid recording of diffractograms at selected core levels (atomic structure). (iv) The ARPES mode reveals the electronic structure within the same probe volume as in steps (ii) and (iii). Measuring steps (iii) and (iv) with circularly polarized light of both helicities yields the CDAD patterns (Circular Dichroism in the Angular Distribution).

In Figure 5, we present a novel application of PED at low kinetic energies, down to the minimum of the inelastic mean free path curve [70]. As in Sec. 3.3, these measurements were performed using the MM at beamline I09 of the Diamond Light Source. Full-field PED patterns with a diameter of ~7 Å$^{-1}$ like those in Fig. 5 are usually acquired in 5 to 10 minutes. Changing photon energy and helicity (for CDAD contrast) is done via an automated routine with small kinetic energy steps down to 2 eV. This high recording speed enables **3D *PED arrays*** I($k_x$,$k_y$,$E_{kin}$) to be captured within a few hours. We have recorded such PED arrays for Si 2*s* and 2*p,* as well as Ge 3*p* and 3*d* core-level photoelectrons, at final state energies ranging from 45 to 400 eV. $E_{final}$ = h$\nu$ - $E_B$ + $V_0$ (with $V_0$ being the inner potential) is the kinetic energy <u>inside</u> the material and defines the electron wavelength. Due to the *gaplens mode's* high angular acceptance, these patterns correspond to polar angle ranges of 0-90° at 45 eV and still 0-30° at 200 eV.

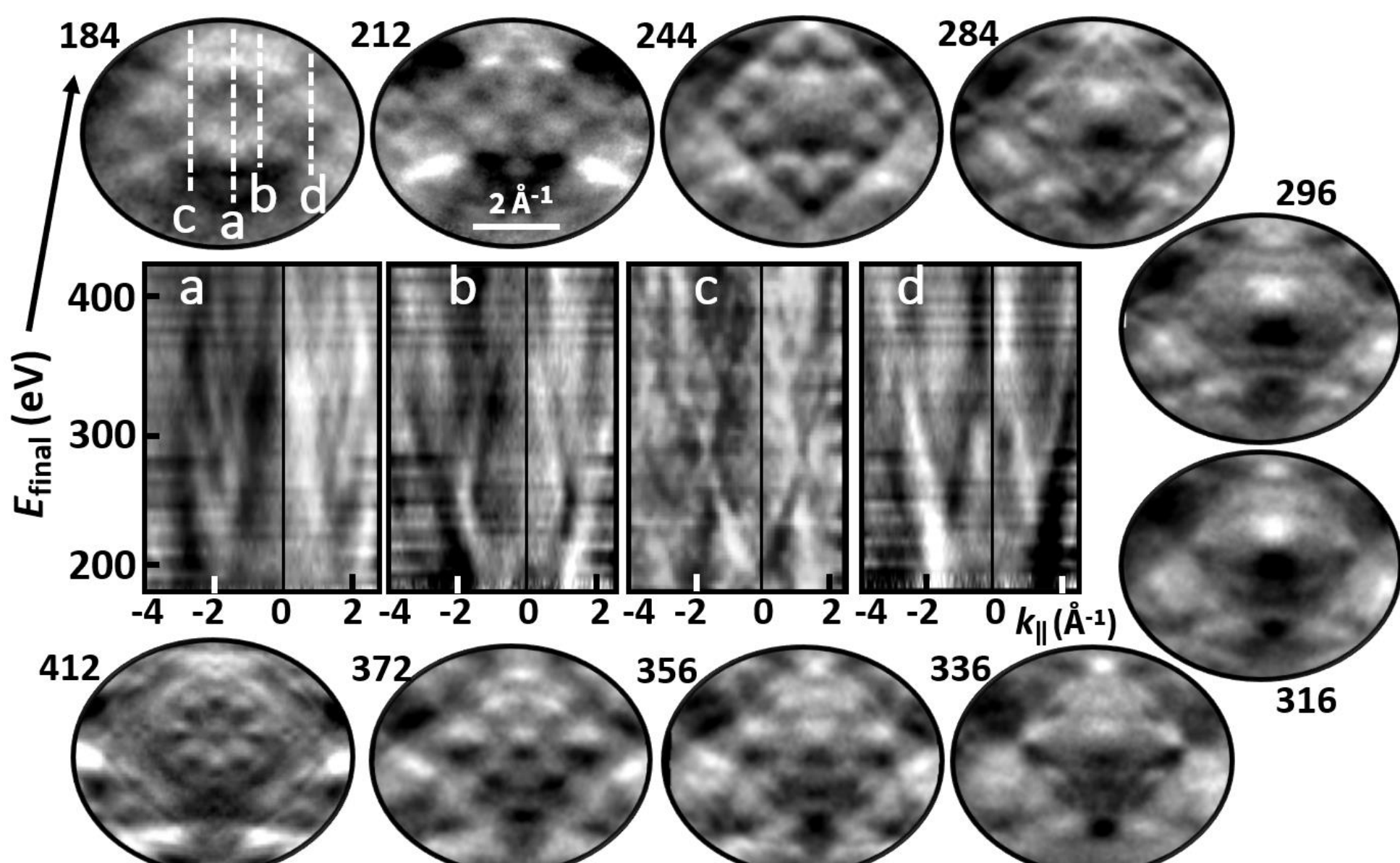

**FIG. 5.** Three-dimensional photoelectron diffraction measurement I($k_x$,$k_y$,$E_{final}$) for Si 2$p_{3/2}$, recorded using the MM at beamline I09 at the Diamond Light Source in a photon energy range from 270 to 500 eV. The measurement was performed in *gaplens mode*, the patterns show CDAD contrast. The outer panels display $k_x$-$k_y$ sections and the inner ones $E_{final}$-*vs*-$k_{\|}$ sections at different positions a-d as marked in the first panel. $E_{final}$ is the kinetic energy inside of the solid.

Fig. 5 shows an example of Si 2$p_{3/2}$ PED patterns in the photon energy range of 270 to 500 eV, corresponding to an $E_{final}$ range of 184 to 412 eV. The figure depicts ten $k_x$-$k_y$ cross sections of the 3D stack at different energies (outer panels) and four $E_{final}$-*vs*-$k_{\parallel}$ sections (inner panels). The image contrast is CDAD, a general phenomenon in core-level photoemission, as demonstrated in Ref. [71]. For the theoretical background, see Refs. [72,73]. At first glance, the abundance of fine details appears to be inconsistent with the low IMFP in this energy range. Understanding these patterns requires new concepts to describe the core-level photoemission process, particularly with regard to the nature of the final state. The Si 2*p* and 2*s,* as well as Ge 3*d* and 3*p* patterns*,* differ significantly even when plotted on the same $E_{final}$ scale [74]. This result emphasizes the importance of the initial orbital angular momentum in the PED process. Calculations are in progress. The gaplens mode was crucial for accessing this data due to the increased contrast and resolution.

### *3.5 ToF-MM in the extreme ultraviolet range at a HHG-based fs photon source*

Several laboratories use ToF-MMs combined with HHG-based XUV sources to study ultrafast, light-induced dynamics using pump-probe techniques [25,26,35–52,75-78]. Most experiments are conducted with cleaved samples, and the pump beam can produce strong nPPE emission. Therefore, the gaplens and repeller modes are of high interest. Photoelectron energies typically range from 16 to 40 eV, with the potential for higher energies up to 100 eV in the future. Setting the cutoff energy for undesired slow electrons to a few eV, close to the signal of interest, poses a greater challenge than in the soft or hard X-ray regions discussed above. Here, we present initial results at $E_{kin}$ = 17 eV. More systematic studies are underway to elucidate the limits of the repeller mode for low-energy electrons.

The first experiments using the multimode front lens at low energies were performed on the instrument at the HHG-based XUV source at the Centre Lasers Intenses et Applications (CELIA) in Bordeaux, France. For details on the experimental setup, see ref. [25]; for scientific results, see refs. [26,76-78]. The standard mode of this setup is the gaplens mode, which produces excellent contrast and resolution. The accelerating converging lens between the sample and the extractor can be tailored using different potentials on the ring electrode and the extractor. This enables operation at a much lower electric field at the sample, typically 3 to 10 times lower than the field in extractor mode. This eliminates the risk of field emission or flashovers for cleaved samples. Additionally, spherical aberration, particularly field curvature, is substantially reduced, and the depth of focus is enhanced. This has been demonstrated in Ref. [21] for field strengths of 880 and 1200 V/mm. The aberration coefficients for both Gaussian and *k*-space imaging are smaller in gaplens mode than in extractor mode. The calculated coefficients for $E_{kin}$ = 100 eV are marked by colored boxes in Tab. 1 of Ref. [21].

We studied a cleaved bulk 2H-$WSe_2$ sample at room temperature using femtosecond XUV pulses (photon energy 21.6 eV, repetition rate 166 kHz, pulse duration ~100 fs). The extractor and ring electrode were set to 8 kV and -2 kV, respectively. This reduced the field strength at the sample surface to 550 V/mm. The equipotential contours of this configuration are shown in Fig. 1(b). Despite the low field strength at the sample, the gap lens is a strong converging lens. With a typical extractor setting of 12 kV for both electrodes, the field strength at the sample would be 1.9 kV/mm, or 3.5 times higher.

Figure 6 shows $k_x$-$k_y$ (a-f) and $E_B$-$k_\parallel$ (g-i) sections, which were measured in *gaplens mode* with $F$ = 550 V/mm. The binding energies of the $k_x$-$k_y$ sections, spanning a 6 eV energy range, are given in the panels. Sections (g) and (h) run along the K-Γ-K′ and K-M-K′ lines, respectively. Section (i) runs parallel to the K-M-K′ line but is shifted slightly toward the Γ-point. A close inspection of section (i) reveals a hybridization gap (avoided crossing), which is marked by an arrow. This gap closes at the M-point, see the arrow in panel (h). The contrast and resolution of this *k*-pattern are high enough to clearly show this local bandgap, which is approximately 100 meV wide. Fig. 6(j) shows an intensity line scan along the dashed line at the K-point in (g), revealing an energy width of 100 meV FWHM. In this room temperature measurement, the energy resolution is dominated by the thermal broadening. A measurement at a sample temperature of 16 K revealed a ToF energy resolution of 29 meV (at a drift energy of 20 eV) and an optical bandwidth of the XUV beam of 33 meV. This corresponds to a Fourier limit of 55 fs, as shown in Fig. 5 of Ref. [25].

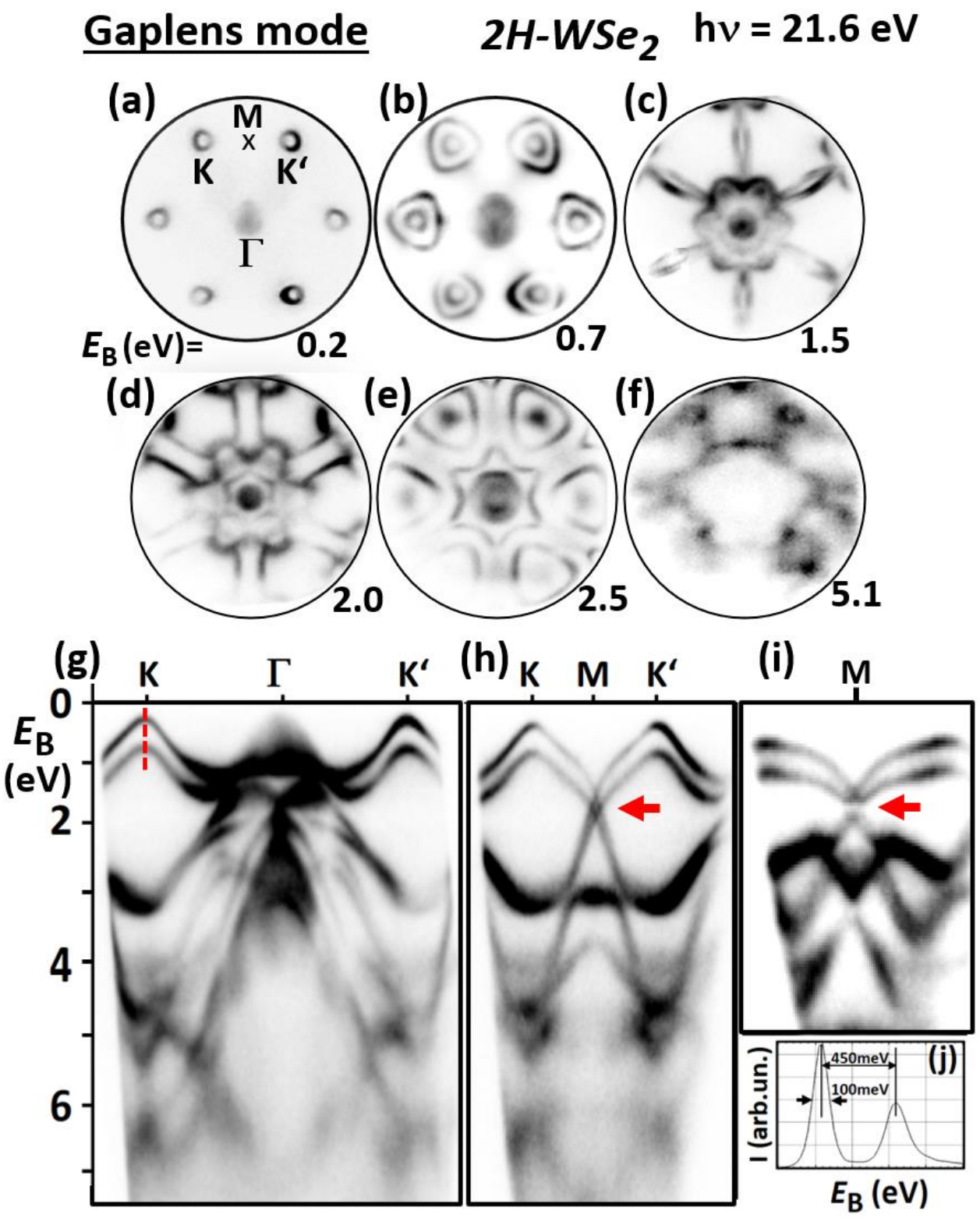

**FIG. 6.** Room-temperature measurement of a 2H-$WSe_2$ sample taken using the ToF-MM at the HHG source at CELIA in Bordeaux at hν = 21.6 eV. The *gaplens mode* ($U_{Ex}$ = 8 kV; $U_R$ = -2 kV) ensured an electric field of only +550 V/mm at the sample surface, as well as higher collection efficiency and lower spherical aberration (field curvature). (a-f) $k_x$-$k_y$ sections at binding energies $E_B$ as stated. (g-i) $E_B$-*vs*-$k_\parallel$ sections along K-Γ-K′, K-M-K′ and parallel to K-M-K′, respectively. The arrows in (h,i) indicate an avoided crossing of the spin-orbit split top band near the M point. (j) Intensity line scan along the dashed line in (g).

Figure 7 shows a measurement of the same sample, but this time using the *repeller mode.* In this case, the field at the sample surface was $F$ = -10 V/mm. This field redirected all electrons with kinetic energies up to approximately 10 eV. The slowest electrons, those below 5 eV, turn around within the first 500 μm above the sample surface. Fig. 1(d) shows the equipotential contours of this mode, where the red contours denote the retarding part of the field. The saddle point $U_S$ of the potential, which defines the cutoff energy, is approximately at -10 V. The $k_x$-$k_y$ section Fig. 7(a) shows that close to the valence-band maximum the $k$-field is reduced to ~50% of the gaplens mode. The $E_B$-*vs*-$k_{\|}$ sections Fig. 7(b) and (c) are shown along the K-Γ and K-M-K' lines, respectively. As the kinetic energy decreases, the horizon shrinks further. As it approaches the cutoff energy of about $E_B$ = 6 eV, the image contracts to a small spot, see Fig. 7(c). With a photon energy of 21.6 eV and an assumed work function of 4.6 eV, electrons at the valence-band maximum have a kinetic energy of 17 eV. Fig. 7(c) shows a cut-off between $E_B$ = 6 and 7 eV, corresponding to $E_{kin}$ = 10 to 11 eV. This confirms the simulated saddle point of $U_S$ = -10V.

Figures 7(d-h) illustrate how the cutoff behavior changes with the repeller field, as seen in the transmitted ToF spectra. At -25 V/mm, only the top valence band can pass, Fig. 7(d). As the field strength decreases, the lower-lying bands appear, Fig. 7(e-g). Finally, with a positive field, the entire spectrum, down to the zero-energy cutoff, is transmitted (h). Since the spectra are plotted on a time-of-flight scale, the energy scales are nonlinear and differ between the panels because the retarding field causes significant time lag.

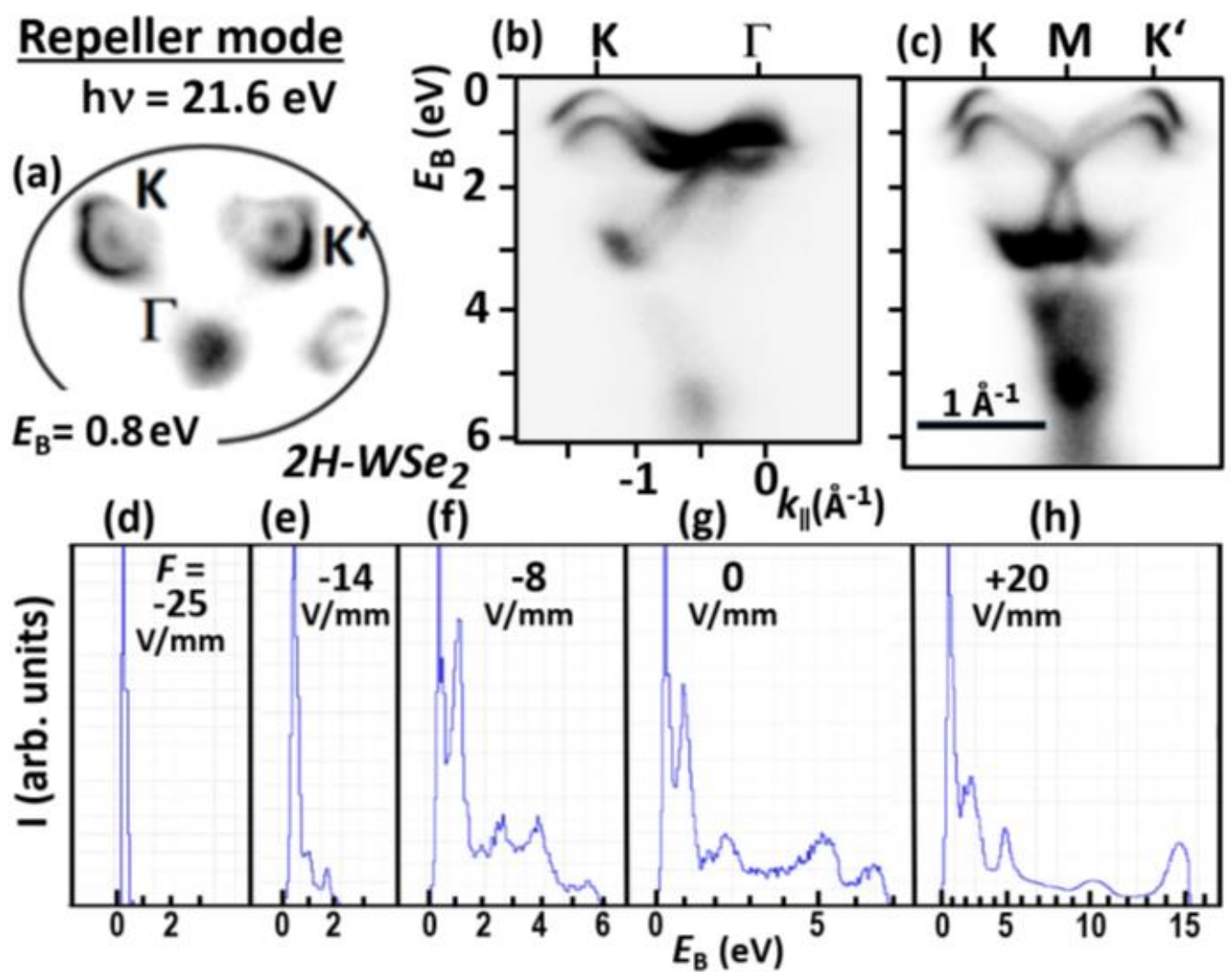

**FIG. 7.** Similar to Fig. 6, but using the *repeller mode* with $U_{Ex}$ = 4.5 kV and $U_R$ = -5.5 kV, yielding $F$ = -10 V/mm at the sample. (a) $k_x$-$k_y$ section at a binding energy of 0.8 eV. (b,c) $E_B$-*vs*-$k_{\|}$ sections along the K-Γ and K-M-K' lines, respectively. (d-h) Sequence of ToF spectra as a function of the field $F$ at the sample surface (values stated in the panels), which is set by varying $U_R$.

At this point, several important comments are in order. First and foremost, a kinetic energy of 17 eV is an extreme case. Simulations show that this energy is close to the limit at which the repeller principle breaks down [21]. Fig. 7 shows the first repeller mode measurement taken at such low energies. Earlier measurements were taken at FLASH with $E_{kin}$= 111.6 eV and $F$ = -21 V/mm [62]. Further systematic studies will elucidate the desired space-charge suppression effect of at such low energies. Eliminating the slow background electrons close to the surface cuts off the main part of the space-charge effect. However, this mode has some drawbacks. Firstly, the $k$-field is significantly reduced, by a factor of two for the low kinetic energies studied in Figs. 6 and 7. Secondly, the usable energy range is also halved. Another drawback is that the low-energy signal from the pump beam is cut off by the repeller voltage. Therefore, the pump-probe overlap must be adjusted using a different mode, such as the gaplens mode, before switching to repeller mode for the $k$-measurement. Whether the repeller mode is beneficial for a specific experiment depends on balancing these disadvantages with the advantage of eliminating space charge effects.

## 4. Performance of the Multimode Lens in Real-Space Imaging

### *4.1 ToF XPEEM - chemical imaging towards 100 nm resolution*

In the previous sections, we have addressed the *k*-imaging modes. Now, we will consider XPEEM, which is real-space imaging at higher energies with chemical selectivity. In XPEEM, the image is recorded using a core-level signal, and contrast is formed by the lateral distribution of selected chemical species. By exploiting the chemical shift, even elements in different chemical states can be distinguished. This differs from threshold PEEM in the UV, where image contrast originates from work function variations or topographic features.

The concept proposed by Tonner [79] was verified by Bauer and coworkers [80] and later extended to higher energies [81,82]. Locatelli et al. [83] and Schmidt et al. [84] found that the spatial resolution of XPEEM at synchrotron sources is ultimately limited by space charge effects. The main contributors to space-charge shifts and broadening are not the interactions of several fast photoelectrons, but rather the Coulomb forces exerted by the large number of secondary electrons pulled into the lens column by the extractor field. The secondary electron background originates from cascade scattering effects and increases significantly at higher photoelectron energies. A ToF-MM experiment at $E_{kin}$ = 1 keV using soft X-ray synchrotron radiation provided a quantitative assessment. At high photon flux, space-charge shifts of >10 eV indicated the presence of a macrocharge of approximately $10^5$ slow electrons trailing each photoelectron. This Coulomb repulsion persists over long periods of time and acts on the photoelectrons over macroscopic distances of tens of millimeters [60].

Using soft X-rays at beamline P04 of PETRA III, we studied the performance of the new front lens in ToF-XPEEM. Our goal was to elucidate the current resolution of element-selective spectral imaging. ToF-MMs can record three-dimensional data arrays $I(x,y,E_{kin})$ in energy intervals of several eV. This allows for rapid mapping of the chemical structure of a sample, including core-level shifts, in one acquisition without scanning. Fig. 8 shows measurements of $WS_2$ flakes with lateral sizes ranging from 3 to 20 µm on an Au substrate. These measurements reveal unique elemental spectroscopic signatures, including chemical shifts that reflect the stoichiometry within each island (see [85] for details). Here, we present results quantifying the lateral resolution and spectroscopic sensitivity of the *gaplens mode*. The measurements were performed with a 300 µm contrast aperture in the back focal plane of the front lens.

The reversal of the chemical contrast is striking when comparing the images recorded at the Au $4f_{7/2}$ and W $4f_{7/2}$ core levels [Figs. 8(a) and (b), respectively]. The former shows the regions of the substrate not covered by flakes, while the latter shows the flakes. Figs. 8(c) and (d) display the corresponding ToF spectra, with the working points for recording the images marked by dashed lines. The spin-orbit splittings are 3.6 eV for Au 4*f* and 2.2 eV for W 4*f*. Fig. 8(e) shows a line scan across to the edge of a flake, marked by the short line in Fig. 8(b). The intensity profile of the raw data reveals an edge onset of 250 nm FWHM.

Earlier work has shown that lateral chromatic aberration, or the change in magnification with varying kinetic energy, can be corrected through data processing [86]. Post-processing the pattern in Fig. 8(b) with this approach and additional machine learning algorithms led to line profiles that were significantly narrower; for details, see [87]. Ray tracing simulations [21] predict resolutions down to 20 nm for the repeller mode, where the space-charge effects are reduced. In threshold PEEM with UV excitation the resolution of the ToF-MM with the new front lens is about 60 nm [25]. Currently, there are no XPEEM data in repeller mode. Based on existing theoretical and experimental results, we conclude that the ToF-XPEEM technique can compete with other instruments, such as the HAXPEEM at PETRA III [82], the Maestro beamline for nanoARPES at the ALS Berkeley [88], the ANTARES beamline at SOLEIL (Saint-Aubin, France) [89] and the aberration-corrected SMART at BESSY [84].

The bottom row of Fig. 8 shows two examples of spectral analysis. Figs. 8(f) and (g) reveal that, compared to the inner part of the flakes, the W $4f_{7/2}$ level is shifted by 0.7 eV to higher binding energy close to the rim of the flakes. Fig. 8(h) shows a markedly different spectral signature of the W $4f_{7/2}$ level, with bright regions appearing in the inner part of the flakes. The corresponding image at the Au $4f_{7/2}$ level, Fig. 8(i), shows that these regions appear dark, indicating that the difference is due to variation in flake thickness. This thickness variation causes the substrate signal to be more strongly absorbed in these regions. The physics of this interesting material will be discussed in a forthcoming paper [85].

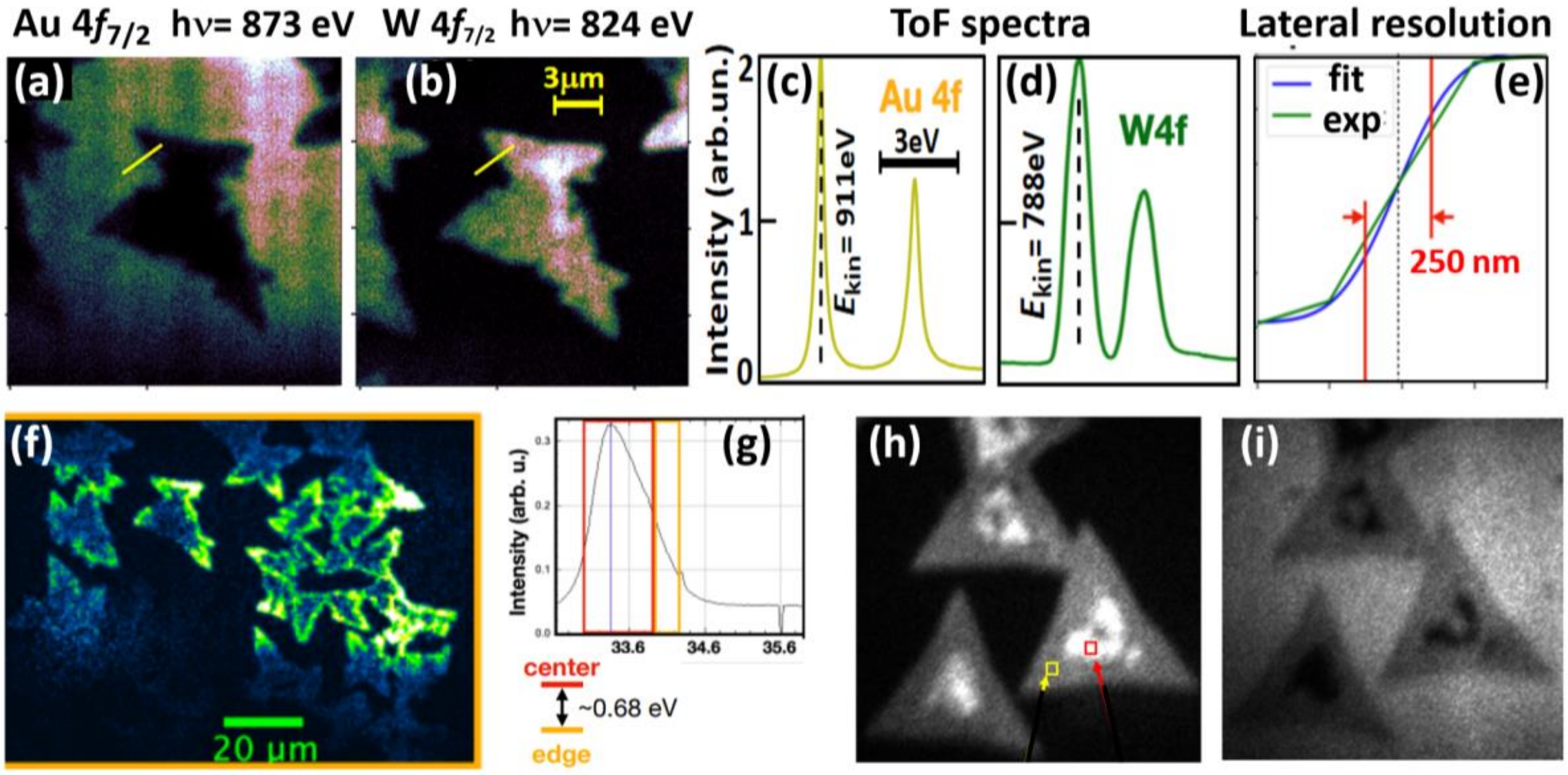

**FIG. 8.** Chemical imaging of $WS_2$ flakes on an Au substrate via ToF-XPEEM in the *gaplens mode*. (a,b) Element-resolved XPEEM images recorded on the Au $4f_{7/2}$ and W $4f_{7/2}$ core-level signals. These images were acquired over a period of 15 minutes each using soft X-rays at beamline P04 of PETRA III at T= 30 K. Different photon energies were used to ensure identical kinetic energies of the signals, allowing the lens settings to remain constant. (c,d) Corresponding ToF spectra; the working points are marked by dashed lines. (e) Measured line-profile across the border of a flake, as marked by the short line in (b). The blue curve is the result of a fit, revealing a width of the edge of 250 nm (FWHM) in the W 4*f* image (raw data). (f) XPEEM image with the energy set to the high-binding-energy wing of the W $4f_{7/2}$ peak, as shown in (g). (h) Inhomogeneous structure in the inner part of some flakes imaged at the W $4f_{7/2}$ peak. (i) The corresponding image at the Au $4f_{7/2}$ peak.

In conclusion, the main advantage of the ToF-XPEEM method is its high count rate, due to capturing energy intervals of several eV in parallel. After the measurement one can scroll through the $I(x,y,E_{kin})$ data array and thus track potential chemical shifts in real space. The flakes were already visible in real time at a frame rate of 1/s when adjusted to the W 4*f* signal. Another unique feature is the ability to switch from real-space to *k*-space imaging. These results are important for future investigations of ultrafast dynamics in real-space imaging [90]. The gaplens mode enables large fields-of-view, limited by the diameter of the extractor hole, see simulation in Fig.6a of Ref. [21]. The PAXRIXS instrument can image converter foils with a diameter of 4 mm. In a prototype lens with a 7 mm extractor hole, we imaged a field of view as large as 6 mm. The repeller and zero-field mode are not suitable for threshold PEEM.

### *4.2 PEEM imaging of a 3D structure – switching off the microlens effect*

Because the 3D potential contour distorts the homogeneous field between the sample and the extractor, PEEM and MM are not ideal for imaging three-dimensional structures. This field deformation reduces resolution and can also cause image artifacts [91,92]. In *gaplens mode,* the reduced field mitigates this effect, which vanishes in *zero-field mode*. We studied this effect using a TEM grid with a 50 μm mesh size. Field penetration into the open meshes causes the 'microlens effect'. See the detailed investigation by Matsuda et al. in Sec. 2.4 of Ref. [93].

Figures 9(a-d) show a series of PEEM images recorded with UV-LED illumination. The local field at the sample was varied from 55 V/mm to 1.5 kV/mm. At the lowest field (a) the grid bars and small, bright impurity spots on the bars are visible without distortion. As the field strength increases, artifact signals appear in the center of the grid meshes [Figs. 9(b and c)], originating from the microlens effect. In the extractor mode, as shown in Fig. 9(d), these artifacts dominate the image, and the bars are almost invisible.

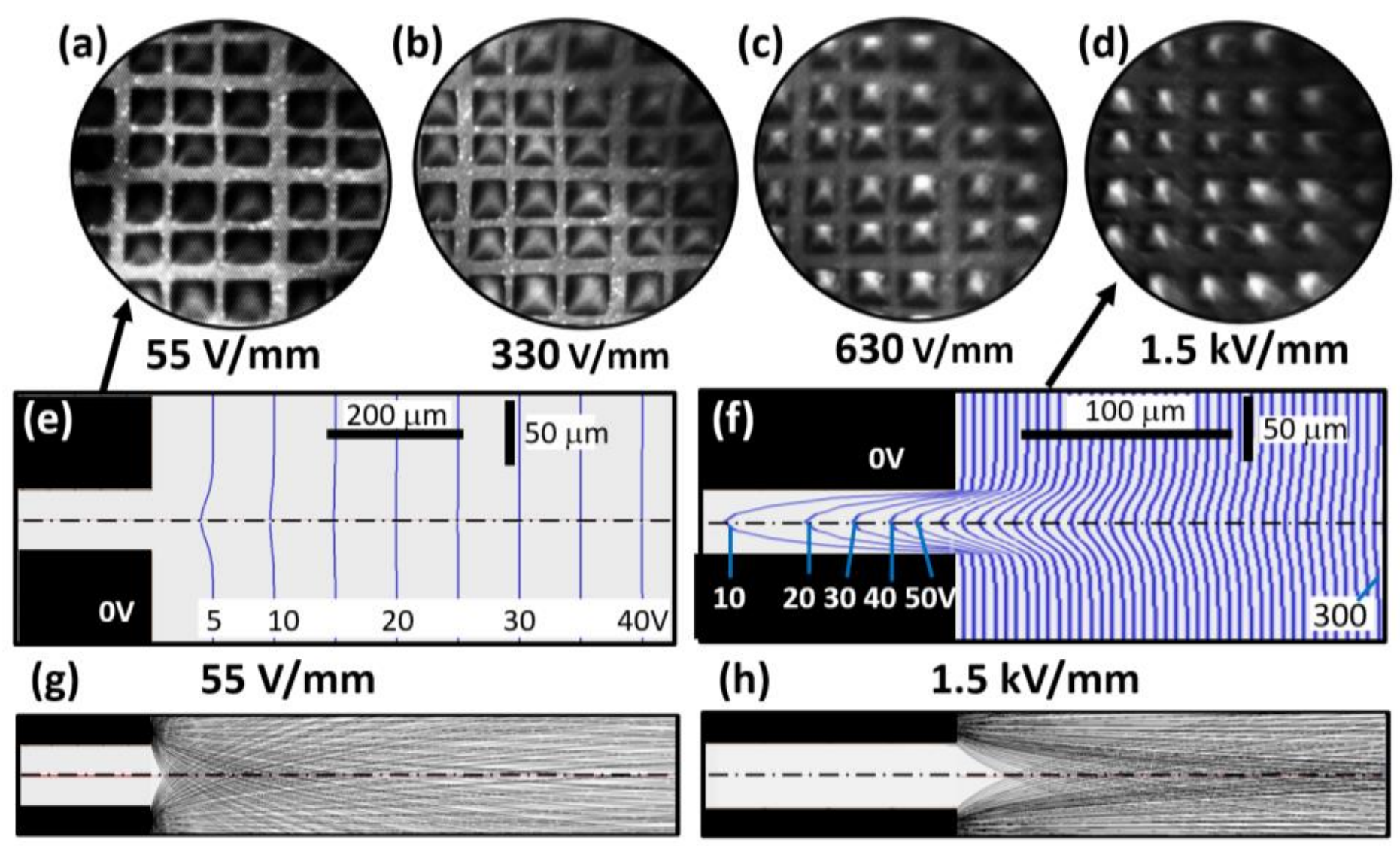

**FIG. 9.** PEEM images of a 3D structure (TEM grid) taken at different field strengths (*F*) at the grid surface. The images were recorded using a UV-LED (265 nm). (a) Image at 55 V/mm; the grid bars (mesh size 50 μm) are visible without distortion. (b, c) As the field strength increases, artifact signals appear in the center of the grid meshes originating from the microlens effect. (d) Image taken in extractor mode ($U_{ex}$ = $U_{rep}$ = 7 kV; *F*= 1.5 kV/mm); the artifacts dominate the images and the grid bars are nearly invisible. (e, f) Equipotential contours (blue) for 55 V/mm and 1.5 kV/mm, respectively. The numbers denote the local potential (sample at 0 V). (g,h) Corresponding electron trajectories. In (h), the microlens effect concentrates the trajectories in front of the mesh hole, leading to the artifacts visible in (d).

Simulations of the equipotential contours (e and f) quantify the microlens effect. In the '*near-zero-field mode'* (e) only the 5 V contour is slightly deformed. However, in *extractor mode*, the field penetrates deeply into the grid meshes, deforming the contours up to large values of the local potential. The corresponding trajectory calculations for electrons with an initial energy of 1 eV are shown in Figs. 9(g,h). There is no visible distortion of the trajectories in (g), while (h) shows that the field distortion at the grid mesh acts as a microlens, concentrating intensity in the center of the meshes. This is consistent with the observation in Fig. 9(d). These data were obtained in UV-PEEM mode, but similar distortion effects occur in momentum measurements of cleaved samples with a 3D structure or operando devices.

## 5. Conclusions and Outlook

Cathode-lens electron microscopes use a strong accelerating field between the sample and the extractor. However, high field strength can cause field emission or space-charge effects, which can be problematic for samples with sharp edges or protrusions resulting from the cleaving process. We have developed a new objective lens that mitigates the risk of field emission or flashovers and space-charge effects. The lens incorporates an additional concentric ring-shaped element around the extractor electrode. Different types of lens field can be created in the gap between the sample and the first electrode by applying different voltages to the extractor and ring electrodes. These potential distributions are associated with different electric fields on the sample surface, including accelerating, zero, and decelerating fields. This multimode lens is therefore beneficial for *k*-space and real-space imaging whenever a high extractor field would cause issues.

The present study aimed to validate electron-optical simulations of the new lens design. A range of applications were studied at different energies, from hard X-rays (up to hν= 6 keV) at the PETRA III synchrotron in Hamburg, to soft X-rays at DIAMOND in Didcot, and even the XUV (hν= 21.6 eV) at the HHG source of CELIA in Bordeaux. In the *gaplens mode* the electric field strength on the sample ranges from 80 V/mm up to several hundred V/mm. By contrast, conventional cathode lenses only achieve useful *k*-resolution at fields of 1000 V/mm or higher. Another benefit of this mode is the increased sample-extractor distance of up to 23 mm. At identical field strengths, the gaplens mode reduces spherical and chromatic aberration compared to the extractor mode. Calculated coefficients are listed in Table 1 of Ref. [21]. Smaller aberrations enable larger fields-of-view in both real-space and momentum imaging.

In the *zero-field* and *repeller modes,* the electric field at the sample either vanishes or decelerates, thereby removing a significant proportion of unwanted slow electrons. For photon energies between 1.4 and 5.8 keV we compared the zero-field mode with the extractor mode (see Figs. 2 and 3). The improved contrast and *k*-resolution are due to the 'switching off' of the primary contribution to the space-charge interaction between photoelectrons and slow background electrons. However, the repeller mode comes at the cost of a smaller *k*-field and a reduced useful energy range due to larger chromatic aberration. We observed a 50% *k*-field reduction at hν= 1488 eV, 86 eV and 21.6 eV (see Figs. 3 and 6). At hard X-ray energies this reduction is smaller (10-20%). The usable energy interval is also reduced by a factor of two, for example from 5 eV to 2.5 eV in Figs. 6g,h,i and 7b,c. Therefore, the usefulness of the repeller mode depends on balancing the reduction in the visible *k*-field and energy width against the advantages of eliminating space-charge artifacts and improving image quality.

*Real-space imaging* also benefits from the new lens modes. For example, in ToF-XPEEM imaging in gaplens mode a line scan of the raw data reveals a resolution of 250 nm (see Fig. 8), which can be enhanced to below 100 nm by data processing [87]. Artifacts caused by the 3D topology of the sample can be eliminated in the near-zero-field mode (see Fig. 9), which counteracts the field distortion caused by the 3D structure that reduces resolution in extractor mode [91,92].

These results will significantly impact future experiments. Momentum microscopes are increasingly being used to study ultrafast phenomena with femtosecond pump-probe techniques (see Refs. [29-52]). In these experiments a large number of slow electrons can be emitted by the pump pulses (infrared or terahertz). Eliminating these electrons using the repeller mode enables much larger pump fluences. For example, in an experiment at FLASH, the pump fluence could be increased by a factor of five after switching to the zero-field mode (see Figs. 3i-n). Our results at $E_{kin}$ = 17 eV (see Figs. 6 and 7) are of central importance to this rapidly developing field of research.

Having a larger working distance between the sample and the first electrode will benefit a variety of experiments. Simulations predict good imaging quality in XPEEM and MM modes at gap sizes of up to 23 mm. Large gap sizes are essential when light optical or gas dosing elements must be placed close to the sample. One forthcoming application requiring a large space between the sample and the first lens electrodes is the PAXRIXS approach, see Fig. 6 in [21] and Refs. [94,95]. In a separate ongoing project, the gap lens captures the real- and *k*-space images of electrons from a gas cell at elevated pressures and temperatures via a small aperture. This enables studying surface chemical reactions in real time. Large gaps and fields close to zero are ideal for future operando experiments involving top electrodes or actuators.

The reduction of the *k*-field in repeller mode will be less severe for ToF-MMs at future HHG-based sources with photon energies in the 100-eV range. Further systematic research is required to determine the method's ultimate limitations at low energies. Future goals also include extending to 4D ($k_x$, $k_y$, $k_z$, $E_{kin}$) recording by exploiting the curvature of the photoemission final-state sphere in *k*-space, as demonstrated in [96]. Lastly, combining spin resolution with high time resolution [97] in ToF-MMs will also benefit from space charge suppression. Eliminating space-charge effects enables higher signal levels and counteracts the significant loss of recording efficiency in spin detectors.

## Acknowledgements

We acknowledge DESY (Hamburg, Germany) and DIAMOND (Didcot, UK), for the provision of beamtimes and experimental facilities. Parts of this research were carried out at PETRA III, beamlines P04 and P22, FLASH I, beamline PG2, and at DIAMOND, beamline I09. Work in Mainz was funded by Bundesministerium für Bildung und Forschung BMBF (projects 05K22UM1, 05K22UM2 and 05K22UM4) and Deutsche Forschungsgemeinschaft DFG through projects SCHO 341/16-1, Transregio SFBs Spin+X TRR175-268565370 (project A02) and Elasto-Q-Mat TRR288–422213477 (project B04). Work in Kiel was funded by the BMBF (project 05K22FK2). Work in Frankfurt was funded by the BMBF (project 05K25RF4) and DFG (project FE 2281/1-1). The instrument at CELIA was funded by the European Union (ERC Starting Grant ERC-2022-STG No.101076639). We further acknowledge funding by the IdEx University of Bordeaux / Grand Research Program "GPR LIGHT", by Quantum Matter Bordeaux, AAP CNRS Tremplin and AAP SMR from Université de Bordeaux.

SF acknowledges funding from the European Union's Horizon Europe research and innovation programme under the Marie Skłodowska-Curie 2024 Postdoctoral Fellowship No 101198277 (TopQMat). J.A.S., P.E.M., D.P., D.L., and Z.X.S. were supported by the U.S. Department of Energy, Office of Basic Energy Sciences, Division of Materials Science and Engineering under contract DE-AC02-76SF00515. J.D.K., G.D. and A.M. were funded by the Linac Coherent Light Source (LCLS), SLAC National Accelerator Laboratory, U.S. Department of Energy, Office of Science, Office of Basic Energy Sciences under Contract No. DE-AC02-76SF00515. Q.N. acknowledges support by the SLAC-Stanford Quantum Initiative Q-FARM Bloch Fellowship and U.S. DOE (DE-AC02–76SF00515). A.K. acknowledges funding provided by the Arnold and Mabel Beckman Foundation, Beckman Young Investigator Award. Sincere thanks go to Andrew J. Mannix, Zhepeng Zhang, Lauren Hoang and Eric Pop (Stanford University, USA) for providing the $WS_2$ sample.

## Data availability

All data shown within this article is available on reasonable request. The authors declare no competing interests.